\documentclass[twocolumn,showpacs,preprintnumbers,amsmath,amssymb,prl]{revtex4}

\usepackage{graphicx}
\usepackage{dcolumn}
\usepackage{bm}

\begin{document}

\title{Shadow on the wall cast by an Abrikosov vortex}

\author{S.~Graser, C.~Iniotakis, T.~Dahm, and N.~Schopohl}

\affiliation{Institut f\"ur Theoretische Physik, Universit\"at T\"ubingen,\\
             Auf der Morgenstelle 14, D-72076 T\"ubingen, Germany}

\date{\today}

\begin{abstract}
At the surface of a $d$-wave superconductor, a zero-energy peak in the quasiparticle 
spectrum can be observed. This peak appears due to Andreev bound states and is maximal if
the nodal direction of the $d$-wave pairing potential is perpendicular to the
boundary. We examine the effect of a single Abrikosov vortex in front of
a reflecting boundary on the zero-energy density of states. We can clearly see a 
splitting of the low-energy peak and therefore a suppression of the zero-energy 
density of states in a shadow-like region extending from the vortex to the boundary. 
This effect is stable for different models of the single Abrikosov vortex, for different 
mean free paths and also for different distances between the vortex center and the
boundary. This observation promises to have also a substantial influence on the
differential conductance and the tunneling characteristics for low excitation energies.
\end{abstract}

\pacs{74.45.+c,74.20.Rp,74.25.-q}

\maketitle

Today, there is common agreement that most high-$T_{c}$ superconductors
exhibit $d$-wave symmetry. An important characteristic of $d$-wave
superconductors is the possible existence of Andreev bound states at its surface 
\cite{Hu,Tanaka,Buchholtz}.
This increase of the local zero-energy quasiparticle density of states
at the surface can clearly be observed in the differential tunneling
conductance as a pronounced zero-bias conductance peak \cite{Iguchi,Lesueur,Covington}.
For specular boundaries, this peak reaches maximum height, if the surface 
is perpendicular to the nodal direction of the $d$-wave. The effect
shrinks, if the orientation is changed \cite{Tanaka,Iguchi}. For an angle of 45 degrees
between the nodal direction and the surface the bound states have vanished
completely. However, it has been pointed out, that for rough surfaces
a similar shape of the zero-bias conductance peak is obtained which is 
independent of the boundary orientation \cite{Fogelstroem}. If a magnetic field 
is applied, the spectral weight of the zero-bias peak decreases and a splitting
of the peak is observed \cite{CovingtonAprili,Aprili,Dagan}.

In this Letter, we study a single Abrikosov vortex 
in front of a specular surface, and we investigate
the effect of the vortex on the local quasiparticle density of states
along the boundary. 
All interesting phenomena of this problem are described within the 
quasiclassical theory, which is valid if 
the coherence length is much larger than the Fermi wavelength.
To calculate the local density of states in the vicinity of the boundary it is
necessary to find numerically stable solutions of the Eilenberger equations
\cite{Eilenberger,Larkin}
that fulfill the appropriate boundary conditions at the specular surface.
For this purpose we use the Riccati parametrization of the quasiclassical
propagator \cite{SchopohlMaki}. Along a trajectory of the kind
${\bm r}(x) = {\bm r}_0 + x \; {\bm v}_F $
the Eilenberger equations of superconductivity reduce to a set of two 
decoupled differential equations of the Riccati type for the functions 
$a(x)$ and $b(x)$
\begin{eqnarray}
\hbar v_F \partial_x  a(x) + \left[ 2 \tilde{\epsilon}_n + \Delta^\dagger a(x) \right] a(x)
- \Delta & = & 0 \nonumber \\
\hbar v_F \partial_x b(x) - \left[ 2 \tilde{\epsilon}_n + \Delta b(x) \right] b(x)
+ \Delta^\dagger & = & 0 \label{Riccati}
\end{eqnarray}
where $i \tilde{\epsilon}_n = i \epsilon_n + {\bm v}_F \cdot \frac{e}{c} 
{\bm A}$ are shifted Matsubara frequencies.
For the simple case of a cylindrical Fermi surface the Fermi velocity can be written as
\begin{equation}
{\bm v}_F = v_F ({\bm e}_1 \cos \theta + {\bm e}_2 \sin \theta)
\end{equation}
The $\theta$- and ${\bm r}$-dependence of the pairing potential $\Delta$ can be factorized 
in the form
\begin{equation}
\Delta ({\bm r}, \theta) = \Delta_0 \chi (\theta) \Psi ({\bm r})
\end{equation}
For a $d_{x^2-y^2}$-wave superconductor the symmetry function takes the form 
$\chi (\theta) = \cos (2 \theta)$ and for the $s$-wave symmetry it becomes constant 
$\chi (\theta) = 1$.
For $\epsilon_n > 0$ the Riccati equations have to be solved using the bulk values as initial 
values at $x = \pm \infty$
\begin{eqnarray}
 a(-\infty) = \frac{\Delta (-\infty)}{\epsilon_n + \sqrt{\epsilon_n^2 + |\Delta
 (-\infty)|^2}} \nonumber \\
 b(+\infty) = \frac{\Delta^\dagger (+\infty)}{\epsilon_n + \sqrt{\epsilon_n^2 + |\Delta
 (+\infty)|^2}}
\end{eqnarray}
The numerical solution of the Riccati equations can be done with minor effort
and leads to rapidly converging results. For the calculation of the local
density of states the imaginary part of the quasiclassical propagator 
has to be integrated over the angle $\theta$ that defines the direction of the 
Fermi velocity. In terms of $a$ and $b$ we have 
\begin{equation}
N({\bm r}_0,E) = \int_0^{2\pi} \frac{d \theta}{2 \pi}  {\rm Re} \;
\left[ \frac{1-a  b}{1+ a b} \right]_{i \epsilon_n \rightarrow E + i \delta}
\label{angularaverage}
\end{equation}
where $E$ denotes the quasiparticle energy with respect to the Fermi level and 
$\delta$ is an effective scattering parameter that corresponds to an inverse 
mean free path. As it is well known, the localized zero-energy state in a 
$d_{x^2-y^2}$-superconductor at a specular 110-boundary arises from the sign 
change in the pairing potential on a quasiparticle trajectory that is reflected 
at the surface. The outgoing particles are Andreev reflected at this potential 
step and interfere with the incident quasiparticles. This interference process 
leads to stable zero-energy trapped states in the vicinity of the boundary, called 
Andreev bound states. The same sign change in the order para\-meter is found on 
a trajectory that passes near the center of a vortex and therefore leads 
to similar localized Andreev bound states inside the vortex core. 
The suppression of the amplitude of the pairing potential around the vortex 
center gives only a small quantitative correction in the calculation of the 
trapped state as we already pointed out before \cite{DGIS}. The influence of the 
boundary for anisotropic superconductors is included within the quasiclassical 
theory if the nonlinear boundary conditions for the quasiparticle propagator 
are applied \cite{Zaitsev, RainerBuchholtz}. 
For the Riccati parametrization a substantial simplification occurs and an 
explicit solution of the nonlinear boundary conditions can be found \cite{Shelankov}.

In the following we assume a totally reflecting surface where
the transparency ${\mathcal T}$ equals zero while the reflectivity ${\mathcal R}$ 
becomes one. In this special case the boundary conditions reduce to
\begin{equation}
a_{\rm out}^{l/r} = a_{\rm in}^{l/r} \hspace{1em} {\rm and} 
\hspace{1em} b_{\rm in}^{l/r} = b_{\rm out}^{l/r}
\end{equation}
In the next step we try to find an appropriate model that describes the pairing 
potential associated with the single vortex in front of the reflecting boundary. 
With the condition that there are no currents flowing across the boundary we 
have to find a phase of the order parameter where the phase gradient vanishes
perpendicular to the boundary. In analogy to the classical boundary value 
problem of electrostatics with a point charge in front of a conducting surface 
we introduce an image vortex on the opposite site of the reflecting boundary.
The pairing potential around a vortex at position ${\bm r}_V$ can be written as 
(see also \cite{DolgovSchopohl})
\begin{equation}
\Psi ({\bm r}) = f(|{\bm r}-{\bm r}_V|) e^{i \Phi({\bm r})}
\end{equation}
The function $f(|{\bm r}-{\bm r}_V|)$ characterizes the amplitude of the pairing potential of the
single vortex. Since we consider a vortex-antivortex pair, the phase $\Phi ({\bm r})$ 
is given as 
\begin{equation}
\Phi ({\bm r}) = {\rm arg} ({\bm r}- {\bm r}_V) - {\rm arg} ({\bm r}- {\bar{\bm r}}_V) 
\end{equation}
The location of the image vortex behind the boundary is defined by
$ {\bar{\bm r}}_V = {\bm r}_V - 2 \hat{\bm n} \langle \hat{\bm n}, {\bm r}_V
\rangle $.
The normal unit vector is given as $\hat{\bm n} = 1/\sqrt{2} ({\bm e}_1 + {\bm e}_2)$ for a 
110-boundary.

In the following, the origin $\mathcal{O}$ of our coordinate system is placed at the boundary 
right between the vortex and the image vortex. The $y$-axis is orientated parallel to 
the boundary. $x_V$ denotes the vortex position on the $x$-axis and also measures the vortex 
to boundary distance. Furthermore it is useful to introduce the coherence length
$\xi = \frac{\hbar v_F}{\Delta_0}$ as the unit of a general length scale. 
We performed calculations of the local density of states for both a model
pairing potential modulus $f(|{\bm r}-{\bm r}_V|)=\tanh (|{\bm r}-{\bm r}_V|/\xi)$ and a 
constant modulus $f(|{\bm r}-{\bm r}_V|)=1$. The latter corresponds to a pure phase vortex. 
The results according to both models show only small quantitative differences. 
In particular, the main qualitative effect we want to present here, the 
shadow on the zero-energy density of states, exists independently. 
Thus, we will restrict our following considerations to the simpler second model. 

\begin{figure}
  \begin{center}
   \includegraphics[width=0.74\columnwidth]{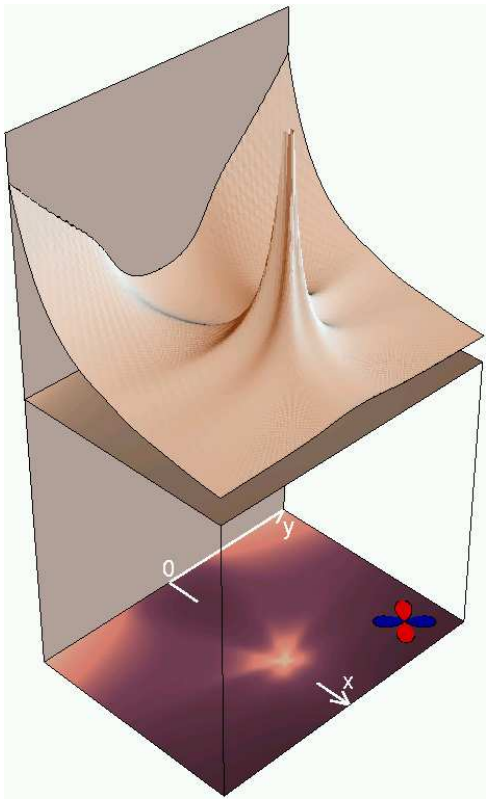}
    \vspace{.05cm}

    \caption{Zero-energy local density of states of a $d_{x^2-y^2}$-wave superconductor. 
    A reflecting 110-boundary is situated at the back of the image, the 
    single Abrikosov vortex can be found at a distance of two coherence lengths $x_V=2 \xi$ in
    front of it. 
    The Andreev bound states at the boundary are seen as the bright regions along the
    $y$-axis, and as the high values in the three-dimensional plot, respectively. 
    The localized states in the vortex center are seen as the peak in the upper image and
    as the bright spot in the image below. One can clearly identify the distinct shadow that 
    emanates from the vortex center to the boundary. An effective scattering parameter 
    of $\delta= 0.1 \Delta_0$ has been used for the calculations. \label{dwaveCompound}}
  \end{center}
\end{figure}

In Fig.~\ref{dwaveCompound} we show the zero-energy local density of states in 
the vicinity of the reflecting boundary. In the upper part of the image, the local density 
of states is displayed as a three-dimensional surface, in the projection below we show 
the same quantity as a density plot. A phase vortex is situated at a distance of two 
coherence lengths $x_V = 2 \xi$ from the surface. 
We assume a $d_{x^2-y^2}$-symmetry of the order parameter and set the boundary
with an angle of 45 degrees to the $b$-axis of the crystal. First we notice, that the 
fourfold symmetry of the local density of states around the vortex is broken. 
Then we also observe a shadow-like suppression of the zero-energy density of 
states in a triangular region between the vortex and the boundary. 

\begin{figure}
  \begin{center}
    \includegraphics[width=0.74\columnwidth]{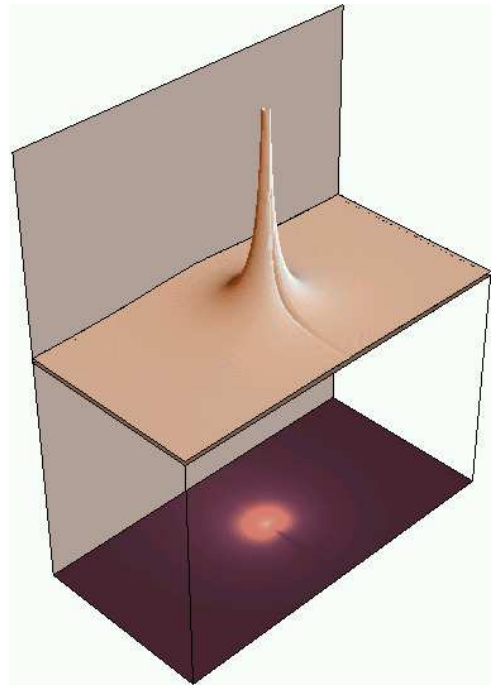}
    \vspace{.05cm}

    \caption{Zero-energy local density of states of a $s$-wave superconductor. 
    Again the reflecting boundary is situated at the back of the image, 
    while the single Abrikosov vortex can be found at a distance of one coherence 
    length $x_V = \xi$ in front of it.
    The high zero-energy density of states of the $s$-wave vortex slightly illuminates 
    the boundary. An effective scattering parameter of $\delta= 0.1 \Delta_0$ 
    has been used for the calculations. \label{swaveCompound} }
  \end{center}
\end{figure}

The picture for an $s$-wave superconductor is totally different. As shown in 
Fig.~\ref{swaveCompound} the vortex has only little influence on the boundary 
density of states. Here, the high zero-energy density of states of the vortex 
rather illuminates the boundary. The small shadow on the right hand side 
of the vortex and the slight line between the vortex and the boundary 
is due to quasiparticles with an inclination angle of 90 degrees, that are trapped 
between the reflecting boundary and the potential step in the vortex core.
Below, we will focus on a $d_{x^2-y^2}$-wave superconductor with an 110-boundary. 
More details of the $s$-wave superconductor and a $d$-wave superconductor with an arbitrary 
orientation of the boundary will be discussed in a following work.

\begin{figure}[t]
  \begin{center}
    \includegraphics[width=1\columnwidth]{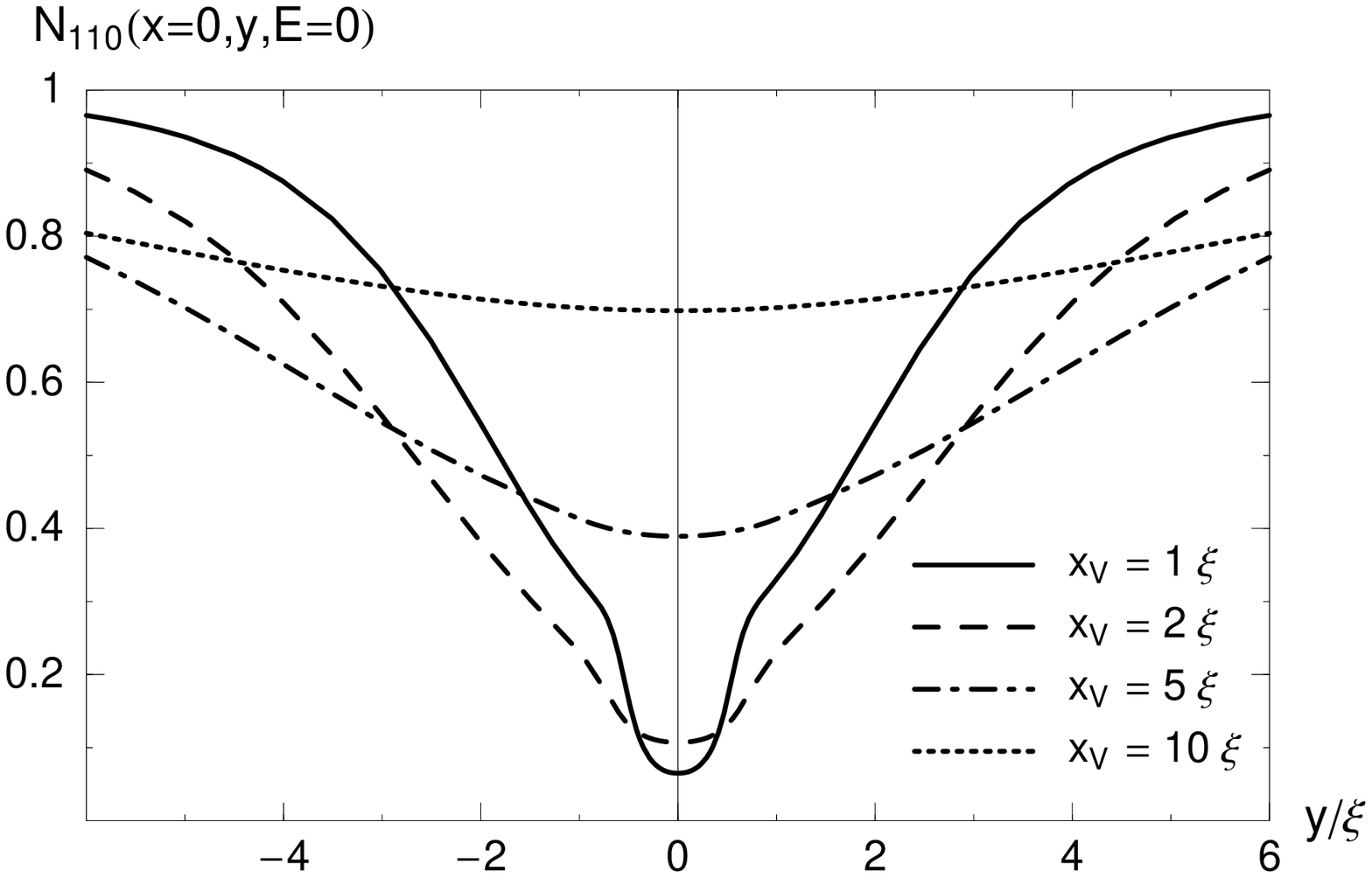}

    \caption{Zero-energy density of states of a $d_{x^2-y^2}$-wave 
    superconductor along a 110-boundary for different vortex positions $x_V$. 
    The curves are normalized to the local density of states 
    at the boundary without vortex ($x_V \rightarrow \infty$). 
    The effective scattering parameter is chosen to be $\delta=0.1 \Delta_0$.
    \label{shadowdepth}}
  \end{center}
\end{figure}

In Fig.~\ref{shadowdepth} we show the zero-energy density of states along the 110-boundary. 
The different curves correspond to different vortex positions $x_V$. The calculations have been 
done using a value of $\delta = 0.1 \Delta_0$. For increasing vortex to boundary
distances $x_V$ we find a decrease of the shadow depth, while the width increases visibly.
Apparently, the shadow effect exists for a wide range of vortex to boundary distances 
$x_V$ even larger than $10 \xi$. For distances smaller 
than one coherence length $\xi$ a selfconsistent calculation of the pairing potential around 
the vortex might become necessary.

\begin{figure}
  \begin{center}
    \includegraphics[width=1\columnwidth]{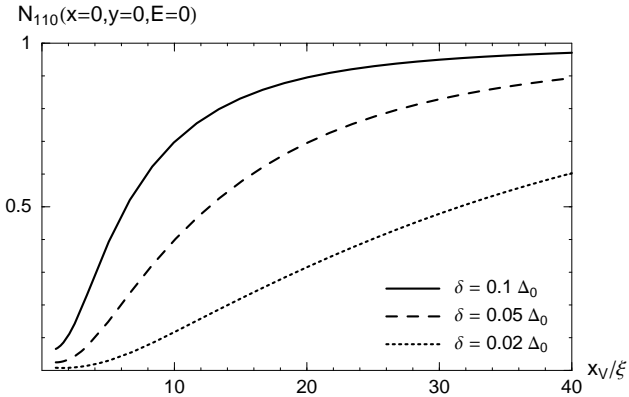}

    \caption{Zero-energy quasiparticle density of states of a $d_{x^2-y^2}$-wave superconductor 
    at the boundary as a function of the vortex position $x_V$. The different curves 
    correspond to different mean free paths related to the inverse of $\delta$. 
    The DOS of each curve is normalized on the corresponding 
    boundary density of states without vortex. 
    \label{ZeDOS_Origin_Distance_del}}
  \end{center}
\end{figure}

In the following we want to obtain a more detailed impression of the vortex shadow at the 
point $\mathcal{O}$. We also want to study the influence of the effective quasiparticle 
scattering parameter $\delta$ introduced in Eq.~(\ref{angularaverage}) on the local 
density of states. 
Without vortex, the value of the zero-energy density of states at the boundary shows a 
sensitive dependence on the decoherence parameter $\delta$. With decreasing $\delta$ the 
value of the zero-energy peak at the boundary increases rapidly. 
In Fig.~\ref{ZeDOS_Origin_Distance_del} we show the zero-energy density of states at the 
point $\mathcal{O}$ as a function of the vortex to boundary distance $x_V$ for different 
values of $\delta$. The curves are normalized to the particular values of the zero-energy 
boundary density of states without a vortex or far away from the vortex center. 
We find that both the range and the relative depth of the shadow increase, if we decrease the 
scattering rate $\delta$. We want to point out, that this shadow effect can be observed in 
clean superconductors with a long mean free path as well as in superconductors with 
higher scattering rates.

In order to explain the suppression of the local zero-energy density
of states at the surface, we now concentrate on a given point in the
shadow region. In order to find the quasiparticle spectrum there, the
angular integration in Eq.~(\ref{angularaverage}) has to be done. For each angle, 
the integrand corresponds to the contribution of a quasiparticle 
trajectory with the direction specified by $\theta $. Due to the phase 
gradient of the order parameter, the energy "seen" by a quasiparticle flying
along a trajectory is shifted. Additionally, this shift itself changes
locally along the trajectory. Thus, for most of the angles, the Riccati-equations
(Eq.~(\ref{Riccati})) are not evaluated at zero-energy. This is
sufficient, however, to miss the sharp zero-energy peak of the bound
state at the surface. As a consequence, the zero-energy density of
states is reduced. In the quasiparticle spectrum
the spectral weight of the bound states is shifted from the Fermi level 
towards higher energies. This effect is similar to the splitting 
of the zero-bias peak due to surface currents \cite{Fogelstroem}.

\begin{figure}[t]
  \begin{center}
    \includegraphics[width=1\columnwidth]{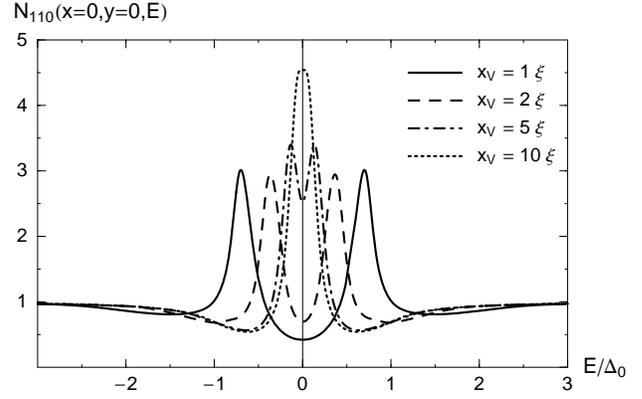}

    \caption{Quasiparticle spectrum of a $d_{x^2-y^2}$-wave superconductor taken at 
     a 110-boundary at the point adjacent to the vortex center ($\mathcal{O}$). 
     The curves are calculated for different vortex distances $x_V$ from the boundary. 
     The effective scattering parameter is chosen to be $\delta=0.1 \Delta_0$.
     \label{LDOSOrigin}}
  \end{center}
\end{figure} 

In Fig.~\ref{LDOSOrigin} we show the local density of states at the 
point $\mathcal{O}$ for different vortex to boundary distances $x_V$ as a 
function of energy. If the Abrikosov vortex is placed in the vicinity of the boundary 
we observe a distinct splitting of the zero-energy
peak. With increasing distance between vortex and boundary the splitting is reduced. 
At $x_V = 10 \xi$ a splitting is no longer visible, while the height of the zero-energy 
peak is still considerably reduced. 

The strong reduction of the zero-energy density of states at the 110-boundary of a 
$d_{x^2-y^2}$-superconductor has of course an important influence on the zero-bias anomaly 
in the tunneling conductance. 
Even in zero magnetic field vortices can remain in the high $T_c$-materials by pinning defects. 
In the vicinity of grain boundaries these pinned vortices will play an important role for the grain 
boundary tunneling due to the reduction of the zero-energy density of states. In a more detailed work we 
will also discuss the influence of a single vortex on the local density of states in the vicinity of a 
rough surface and we will consider several interesting boundary geometries apart from the flat surface.

\acknowledgments

S.~G. is supported by the 'Graduiertenf\"orderungs\-programm des Landes
Baden-W\"urttemberg'. C.~I. is grateful to the German National Academic Foundation. 
Part of this work was funded by the 'Forschungsschwerpunkt "Quasiteilchen"
des Landes Baden-W\"urttemberg'.


\begin{thebibliography}{***}


\bibitem{Hu} C.~R.~Hu, 
{Phys.~Rev.~Lett.} {\bf 72}, 1526 (1994). 

\bibitem{Tanaka} Y.~Tanaka and S.~Kashiwaya, {Phys.~Rev.~Lett.} {\bf 74}, 3451 
(1995). 

\bibitem{Buchholtz} L.~J.~Buchholtz, M.~Palumbo, D.~Rainer, and J.~A.~Sauls, 
{J.~Low~Temp.~Phys.} {\bf 101}, 1099 (1995).

\bibitem{Iguchi} I.~Iguchi, W.~Wang, M.~Yamazaki, Y.~Tanaka, and S.~Kashiwaya,
{Phys.~Rev.~B} {\bf 62}, R6131 (2000).

\bibitem{Lesueur} J.~Lesueur, L.~H.~Greene, W.~L.~Feldmann, and A.~Inam,
{Physica C} {\bf 191}, 325 (1992).

\bibitem{Covington} M.~Covington, R.~Scheuerer, K.~Bloom, and L.~H.~Greene,
{Appl.~Phys.~Lett.} {\bf 68}, 1717 (1996).

\bibitem{Fogelstroem} M.~Fogelstr\"om, D.~Rainer, and J.~A.~Sauls,
{Phys.~Rev.~Lett.} {\bf 79}, 281 (1997).

\bibitem{CovingtonAprili} M.~Covington, M.~Aprili, E.~Paraoanu, L.~H.~Greene,
F.~Xu, J.~Zhu, and C.~A.~Mirkin,
{Phys.~Rev.~Lett.} {\bf 79}, 277 (1997).

\bibitem{Aprili} M.~Aprili, E.~Badica, and L.~H.~Greene,
{Phys.~Rev.~Lett.} {\bf 83}, 4630 (1999).

\bibitem{Dagan} Y.~Dagan and G.~Deutscher,
{Phys.~Rev.~Lett.} {\bf 87}, 177004 (2001).

\bibitem{Eilenberger} G.~Eilenberger, 
{Z.~Phys.} {\bf 214}, 195 (1968). 

\bibitem{Larkin} A.~I.~Larkin and Yu.~N.~Ovchinnikov, 
{Zh.~Eksp.~Teor.~Fiz.} {\bf 55}, 2262 (1968) [{Sov.~Phys.~JETP} {\bf 28}, 1200
(1969)].  

\bibitem{SchopohlMaki} N.~Schopohl and K.~Maki, 
{Phys.~Rev.~B} {\bf 52}, 490 (1995);
N.~Schopohl, cond-mat/9804064 (unpublished).

\bibitem{DGIS} T.~Dahm, S.~Graser, C.~Iniotakis, and N.~Schopohl, 
{Phys.~Rev.~B} {\bf 66}, 144515 (2002).

\bibitem{RainerBuchholtz} L.~J.~Buchholtz and D.~Rainer, 
{Z.~Phys.~B} {\bf 35}, 151 (1979).

\bibitem{Zaitsev} A.~V.~Zaitsev, 
{Zh.~Eksp.~Teor.~Fiz.} {\bf 86}, 1742 (1984) [{Sov.~Phys.~JETP} 
{\bf 59}, 1015 (1984)].

\bibitem{Shelankov} A.~Shelankov and M.~Ozana,
{Phys.~Rev.~B} {\bf 61}, 7077 (2000).

\bibitem{DolgovSchopohl} O.~V.~Dolgov and N.~Schopohl, 
{Phys.~Rev.~B} {\bf 61}, 12389 (2000). 

\end{thebibliography}
\end{document}